\documentclass{article}
\usepackage{graphicx}  % got figures? uncomment this
\title{A de Broglie-Bohm Like Model for Klein-Gordon Equation\\submited to Nuovo Cimento}
\author{O. Chavoya-Aceves\\
Camelback H. S., Phoenix, Arizona, USA.\\
chavoyao@yahoo.com}

\newcommand{\grad}{\vec{\nabla}}

\begin{document}

\maketitle

\begin{abstract}
A de Broglie-Bohm like model of Klein-Gordon equation, that leads
to the correct Schr\"odinger equation in the low-speed limit, is
presented. Under this theoretical framework, that affords an
interpretation of the \emph{quantum potential}, the main
assumption of the de Broglie-Bohm interpretation---that the local
momentum of particles is given by the gradient of the phase of the
wave function---is not but approximately correct. Also, the number
of particles is not locally conserved. Furthermore, the
representation of physical systems through wave functions wont be
complete.

\end{abstract}

\section{Introduction}
It is generally accepted that the Klein-Gordon equation has not a
consistent de Broglie-Bohm like interpretation\cite[p.
498-503]{HOLLAND}\cite{BOHM}. However, in this paper we show that
through the introduction of a hidden variable $\Phi$ such that:
\[
p_\mu=-\partial_\mu S - \partial_\mu \Phi,
\]
where $S$ is the phase of the wave function, a solution to this
problem can be easily found, in such way that it leads to the
correct Schr\"odinger equation in the low-speed limit, and affords
a sound interpretation of the quantum potential.

Within the theoretical framework of this work, the main assumption
of the de Broglie-Bohm interpretation of quantum mechanics is not
valid and a picture of particles moving under the action of
electromagnetic field alone is recovered. However, the number of
particles is not locally conserved. Also, given that a
\emph{hidden variable} has been introduced---on the grounds of
some general electrodynamic considerations, presented in the first
section---the representation of physical systems through wave
functions wont be complete and, therefore, as foreseen by
Einstein, Podolsky, and Rosen, quantum mechanics wont be a
complete theory of motion.

\section{On the Motion of Particles Under the Action of an Electromagnetic Field}

Consider a non-stochastic ensemble of particles whose motion is
described by means of a function
\[
\vec{r}=\vec{r}(\vec{x},t),
\]
such that $\vec{r}(\vec{x},t)$ represents the position, at time
$t$, of the particle that passes through the point $\vec{x}$ at
time \emph{zero}---the $\vec{x}$-particle. This representation of
motion coincides with the lagrangian representation used in
hydrodynamics\cite{ARIS}.

As to the function $\vec{r}(\vec{x},t)$, we suppose that it is
invertible for any value of $t$. In other words, that there is a
function $\vec{x}=\vec{x}(\vec{r},t)$, that gives us the
coordinates, at time zero, of the particle that passes through the
point $\vec{r}$ at time $t$. Also, we suppose that
$\vec{r}(\vec{x},t)$ is a continuous function, altogether with its
derivatives of as higher order as needed to secure the validity of
our conclusions.

According to the definition of $\vec{r}(\vec{x},t)$, the velocity
of the $\vec{x}$-particle at time $t$ is
\[
\vec{u}(\vec{x},t)=\left( \frac{\partial \vec{r}}{\partial
t}\right)_{\vec{x}}.
\]

Writing $\vec{x}$ as a function of $\vec{r}$ and $t$, we get the
velocity field at time $t$:
\[
\vec{v}(\vec{r},t)=\vec{u}(\vec{x}(\vec{r},t),t).
\]

As it is known from fluid kinematics, the corresponding field of
acceleration is:
\[
\frac{\partial \vec{v}}{\partial t}+ (\vec{v}\cdot \grad)\vec{v}.
\]

From this, considering that
\[
(\vec{v}\cdot \vec{\nabla})\vec{v} = \grad
v^2/2-\vec{v}\times(\grad\times\vec{v})
\]
we can write the corresponding field of force:
\[
\vec{f}=m\left( \vec{e}+\frac{1}{c}\vec{v}\times\vec{h}\right)
\]
where
\begin{equation}\label{analog of electric field}
\vec{e}=-\frac{1}{c}\frac{\partial \vec{a}}{\partial t}-\grad a_0,
\end{equation}
\begin{equation}\label{analog of magnetic field}
  \vec{h}=\grad\times\vec{a},
\end{equation}
\begin{equation}\label{analog of scalar potential}
  a_0=-c^2-v^2/2,
\end{equation}
\begin{equation}\label{analog of vector potential}
  \vec{a}=-c\vec{v},
\end{equation}
and $c$ is a constant, with dimensions of speed, that we shall
take equal to the velocity of light.

Equations (\ref{analog of electric field}) and (\ref{analog of
magnetic field}) are formally analogous to the definition of the
electric and magnetic fields from the electrodynamic potentials.
From them we can prove that
\[
\grad\cdot\vec{h}=0,
\]
and
\[
\grad\times\vec{e}=-\frac{1}{c}\frac{\partial \vec{h}}{\partial
t},
\]
analogous to the first pair of Maxwell equations. Furthermore, the
right members of equations (\ref{analog of scalar potential}) and
(\ref{analog of vector potential}), but for a constant factor,
correspond to the low-speed approximation of the components of the
relativistic four-velocity, which suggest us to investigate the
properties of this field, defined as:
\[
v^\mu=\left(\frac{c}{\sqrt{1-v^2/c^2}},\frac{\vec{v}}{\sqrt{1-v^2/c^2}}\right)
\]
Using four-dimensional tensorial notation, the derivative of
$v_\mu$ with respect to the proper time, along the world-line of
the corresponding particle is:
\[
\frac{d v_\mu}{ds}=v^\nu \partial_\nu v_\mu
\]
From the identity
\begin{equation}\label{square of four-velocity}
  v^\nu v_\nu=c^2
\end{equation}
we prove that
\[
v^\nu\partial_\mu v_\nu=0,
\]
that allows us to write
\begin{equation}\label{four acceleration}
  \frac{d v_\mu}{ds}=(\partial_\nu v_\mu -\partial_\mu
  v_\nu)v_\nu.
\end{equation}

If the four-force acting on the ensemble of particles has an
electromagnetic origin, it is given by the expression
\begin{equation}\label{lorentz}
  f_\mu=\frac{q}{c}F_{\mu\nu}v^\nu,
\end{equation}
where
\begin{equation}\label{faraday}
  F_{\mu\nu}=\partial_\mu A_\nu-\partial_\nu A_\mu
\end{equation}
is Faraday's tensor, and $A_\mu$ is the electrodynamic
four-potential.

From the law of motion
\[
m\frac{dv_\mu}{ds}=f_\mu,
\]
and equations (\ref{four acceleration}) to (\ref{faraday}) we
prove that
\begin{equation}\label{fundamental equation}
  (\partial_\mu p_\nu - \partial_\nu p_\mu)v^\nu=0,
\end{equation}
where
\begin{equation}\label{four-momentum}
  p_\mu=m v_\mu + \frac{q}{c}A_\mu.
\end{equation}

There is a class of solutions of equations (\ref{fundamental
equation}) where $p_\mu$ is the four-gradient of a function of
space-time coordinates
\begin{equation}\label{definition of mechanical action}
  p_\mu=-\partial_\mu \phi.
\end{equation}
Therefore,
\[
\partial_\mu p_\nu -\partial_\nu p_\mu =0,
\]
and (\ref{fundamental equation}) is obviously satisfied.

Equation (\ref{four-momentum}) can be written in the form
\begin{equation}\label{kinetic momentum}
  m v_\mu=-\partial_\mu \phi -\frac{q}{c} A_\mu
\end{equation}

If the four-potential meets the Lorentz condition,
\begin{equation}\label{lorentz condition}
  \partial^\mu A_\mu =0,
\end{equation}
equation (\ref{kinetic momentum}) is analogous to the
decomposition of a classical, three-dimensional field, into an
irrotational and a solenoidal part. The electromagnetic field
appears thus as determining the four-dimensional vorticity of the
field of \emph{kinetic momentum} $m v_\mu$.

Equation (\ref{square of four-velocity}) tells us that functions
$\phi$ are not arbitrary, but are subject to the condition:
\begin{equation}\label{relativistic hamilton-jacobi}
  (\partial^\mu \phi
  +\frac{q}{c}A^\mu)(\partial_\mu\phi+\frac{q}{c}A_\mu)=m^2 c^2,
\end{equation}
which is the relativistic Hamilton-Jacobi equation.\cite[Ch.
VIII]{LANCZOS}

The components of the kinetic momentum are:
\begin{equation}\label{components of kinetic momentum}
  \left(\frac{K}{c},\vec{p}\right)=\left(-\frac{1}{c}\frac{\partial \phi}{\partial t}-\frac{q}{c}V,\grad \phi -
  \frac{q}{c}\vec{A}\right),
\end{equation}
where $V$ and $\vec{A}$ are the components of the electrodynamic
four-potential.

In the low-speed limit, we have
\[
K\approx mc^2 + \frac{p^2}{2m}.
\]
From this and (\ref{components of kinetic momentum}) we get:
\begin{equation}\label{non-relativistic hamilton-jacobi}
  \frac{\partial \phi}{\partial t}+ \frac{(\grad \phi -
  \frac{q}{c}\vec{A})^2}{2m}+ q V + m c^2 =0,
\end{equation}
that, but for the constant term $m c^2$, is the non-relativistic
Hamilton-Jacobi equation.

Notice that
\[\oint mv_\mu dx^\mu = -\frac{q}{c}\oint A_\mu dx^\mu,\]
for any closed path in space time. In particular
\[\oint\vec{p}\cdot d^3
\vec{x}=-\frac{q}{c}\oint\vec{A}\cdot d^3\vec{x},\] for any closed
path of simultaneous events in three-dimensional space.
\section{A de Broglie-Bohm Like Interpretation of Klein-Gordon Equation}
Consider the Klein-Gordon equation describing an ensemble of
spin-less particles with mass $m$ and charge $q$:
\[
(i\hbar \partial^\mu -\frac{q}{c}A^\mu)(i\hbar \partial_\mu
  -\frac{q}{c}A_\mu)\Psi=0,
\] where $A^\mu$ satisfies the Lorentz condition (\ref{lorentz
condition}). In other words:
\begin{equation}\label{klein gordon}
  -\hbar^2\partial^\mu\partial_\mu\Psi
  -\frac{2i\hbar}{c}A^\mu\partial_\mu\Psi+\frac{q^2}{c^2}A^\mu
  A_\mu\Psi = m^2c^2 \Psi.
\end{equation}

We can easily show that
\begin{equation}\label{first continuity equation}
  \partial^\mu \left(\frac{i\hbar}{2}(\Psi^\star\partial_\mu \Psi-\Psi\partial_\mu
  \Psi^\star)-\frac{q}{c}A_\mu\Psi^\star\Psi\right)=0.
\end{equation}

By the Madelung substitution
\begin{equation}\label{madelung substitution}
  \Psi =\sqrt{\rho}e^{\frac{i}{\hbar}S},
\end{equation}
equation (\ref{second continuity equation}) is transformed into
\begin{equation}\label{second continuity equation}
  \partial^\mu(\rho (-\partial_\mu S-\frac{e}{c}A_\mu))=0,
\end{equation}
that, in view of (\ref{kinetic momentum}), tells us that
\begin{equation}\label{density of particles}
  \rho=\Psi^\star \Psi,
\end{equation}
and
\begin{equation}\label{wrong fourvelocity}
  w_\mu=-\frac{\partial_\mu S}{m}-\frac{q}{mc}A_\mu,
\end{equation}
might be interpreted as the density of particles in the (local)
system of reference, where the particles are at rest, and the
field of four-velocity, respectively. (Hence $p_\mu=-\partial_\mu
S$.) This interpretation is not sound, however, because $w_\mu$ is
not an implicitly time-like four-vector. Actually, from
(\ref{klein gordon}) and (\ref{madelung substitution}), it can be
shown that
\begin{equation}\label{introduction of quantum potential}
(-\partial^\mu S
  -\frac{q}{c}A^\mu)  (-\partial_\mu S -\frac{q}{c}A_\mu) =
  m^2c^2+\hbar^2 \frac{\partial^\mu\partial_\mu
  \sqrt{\rho}}{\sqrt{\rho}},
\end{equation}
in plain disagreement with (\ref{relativistic hamilton-jacobi}).

However, considering that $w_\mu$ has already the required
four-vorticity, we'll suppose that there is a function $\Phi$ such
that
\begin{equation}\label{klein-gordon four-velocity}
  m v_\mu = -\partial_\mu S -\partial_\mu \Phi - \frac{q}{c}A_\mu.
\end{equation}
From (\ref{relativistic hamilton-jacobi}) and (\ref{introduction
of quantum potential}) we can show that $\Phi$ has to satisfy the
condition
\begin{equation}\label{condition for Phi}
  2mv^\mu \partial_\mu \Phi +\partial^\mu\Phi\partial_\mu \Phi=
 \hbar^2 \frac{\partial^\mu\partial_\mu \sqrt{\rho}}{\sqrt{\rho}}.
\end{equation}

We'll show that this is a sound hypothesis by proving that it
leads to the correct equations in the low-speed limit, where we
suppose that
\[
\frac{1}{c^2}\left(\frac{\partial \Phi}{\partial t}\right)^2= 0,\
\  \ \ \ \ \frac{\hbar^2}{c^2\sqrt{\rho}}\frac{\partial^2
\sqrt{\rho}}{\partial t^2}= 0,
\]
in such way that (\ref{condition for Phi}) can be rewritten, using
classical, three-dimensional, notation, as
\begin{equation}\label{first classical approximation}
2m\frac{\partial \Phi}{\partial t}+ 2m\vec{v}\cdot \grad\Phi
-(\grad{\Phi})^2=-\hbar^2\frac{\Delta \sqrt{\rho}}{\sqrt{\rho}}
\end{equation}
or,
\begin{equation}\label{second classical approximation}
\frac{\partial \Phi}{\partial t}+(\vec{v}-\grad \Phi)\cdot
\grad\Phi
+\frac{(\grad\Phi)^2}{2m}=-\frac{\hbar^2}{2m}\frac{\Delta
\sqrt{\rho}}{\sqrt{\rho}}
\end{equation}

In other words:
\begin{equation}\label{third classical approximation}
  \frac{(\nabla S -\frac{q}{c}\vec{A})}{m}\cdot \nabla \Phi+\frac{(\grad\Phi)^2}{2m} =-\frac{\hbar^2}{2m} \frac{\Delta
  \sqrt{\rho}}{\sqrt{\rho}}-\frac{\partial \Phi}{\partial t}
\end{equation}

From this we can build a low-speed limit approximation of the
kinetic energy:
\begin{equation}\label{fourth classical approximation}
  K \approx m c^2 + \frac{(\nabla S + \nabla \Phi  - \frac{q}{c}\vec{A})^2}{2m}%
   = m c^2 +\frac{(\nabla S -\frac{q}{c} \vec{A})^2}{2m}-\frac{\hbar^2}{2m} \frac{\Delta
  \sqrt{\rho}}{\sqrt{\rho}}-\frac{\partial \Phi}{\partial t},
\end{equation}

Finally, given that in this case,
\begin{equation}\label{klein gordon kinetic energy}
  K=-\frac{\partial (S+\Phi)}{\partial t}-qV.
\end{equation}
we conclude that, in the low-speed limit:
\begin{equation}\label{classical hamilton jacobi}
\frac{\partial S}{\partial t}+m c^2+\frac{(\nabla S -\frac{q}{c}
  \vec{A})^2}{2m}-\frac{\hbar^2}{2m} \frac{\Delta
  \sqrt{\rho}}{\sqrt{\rho}}+qV= 0,
\end{equation}
that, but for the constant term $mc^2$, is the classical
Hamilton-Jacobi equation with quantum potential. From this and
(\ref{second continuity equation}) the corresponding Schr\"odinger
equation is easily recovered.

In the general case, we get a picture of classical particles
moving under the action of the electromagnetic field alone: there
is not \emph{quantum potential}. The field of four-velocity is
given by (\ref{klein-gordon four-velocity}), where $S$ is the
phase of the wave function and $\Phi$ is a \emph{hiden variable}.
$\rho$, $S$ and $\Phi$ are solutions of a system of non-linear
differential equations: (\ref{second continuity equation}),
(\ref{introduction of quantum potential}), and
\begin{equation}\label{hidden equation}
  (-\partial^\mu S -\partial^\mu \Phi -\frac{q}{c}A^\mu)  (-\partial_\mu S -\partial_\mu \Phi
  -\frac{q}{c}A_\mu)=m^2 c^2,
\end{equation}
that, however, does not determine them completely. But for the
restrictions imposed in the low-speed limit, the function $\Phi$
could be equal to $\phi-S$ for any solution of equation
(\ref{relativistic hamilton-jacobi}).

The number of particles is not locally conserved. From
(\ref{second continuity equation}) and (\ref{klein-gordon
four-velocity}) we get:
\begin{equation}\label{third continuity equation}
  \partial^\mu(\rho
  v_\mu)=-\frac{\partial^\mu(\rho\partial_\mu\Phi)}{m}.
\end{equation}

The $\Phi$ function appears thus related to local
creation/annihilation processes, which leads to the appearance of
a \emph{quantum force} in the corresponding hydrodynamic model.

Let
\begin{equation}\label{energy tensor}
  T_{\mu\nu}= m \rho v_\mu v_\nu
\end{equation}
Then
\begin{equation}\label{euler equation}
\partial^\nu T_{\mu\nu}=m\rho v_\nu \partial^\nu v_\mu+ mv_\mu
\partial^\nu(\rho v_\nu)=\frac{q}{c}\rho v^\nu F_{\mu\nu}-v_\mu
\partial^\nu(\rho \partial_\nu \Phi)
\end{equation}

Equations (\ref{energy tensor}) and (\ref{euler equation})
describe a pressure-less fluid of charged particles under the
action of the electromagnetic fluid and a force:
\begin{equation}\label{quantum force}
K_\mu= -v_\mu
\partial^\nu(\rho \partial_\nu \Phi)
\end{equation}
associated---through (\ref{third continuity equation})---to local
creation/annihilation processes.\cite{EINSTEIN}
\section{The Low-Speed Limit Revisited}
In the low speed limit, the equations of quantum mechanics for
spin-less particles can be proved to be equivalent to:
\begin{equation}\label{first madelung fluid}
 m\left(\frac{\partial \vec{u}}{\partial t}+(\vec{u}\cdot
  \nabla)\vec{u}\right)=q\textbf{E}+\frac{q}{c}\vec{u}\times\textbf{H}-\grad Q
\end{equation}
and
\begin{equation}\label{second madelung fluid}
  \frac{\partial \rho}{\partial t}+\grad\cdot(\rho \vec{u})=0,
\end{equation}
where
\begin{equation}\label{densidad}
  \rho=\Psi^\star \Psi,
\end{equation}
\begin{equation}\label{first component of velocity}
  m\vec{u}=\grad S -\frac{q}{c}\vec{A},
\end{equation}
and
\begin{equation}\label{quantum potential}
  Q=-\frac{\hbar^2}{2m}\frac{\Delta \sqrt{\rho}}{\sqrt{\rho}},
\end{equation}
is the \emph{quantum potential}.

From (\ref{first classical approximation})
\[
-\grad Q = -\frac{\partial \grad \Phi}{\partial
t}-(\vec{v}\cdot\grad)\grad\phi-(\grad\phi\cdot\grad)\vec{v}-\grad\phi\times(\grad\times\vec{v})+\grad
\frac{(\grad{\Phi})^2}{2m}
\]
On this basis, considering that in this limit the field of
velocity is given by:
\begin{equation}\label{definition of velocity}
  m\vec{v}=\grad S + \grad{\Phi}-\frac{q}{c}\vec{A},
\end{equation}
and that, therefore,
\[
m\vec{v}=m\vec{u}+\grad \Phi,
\]
and
\[
m\grad \times\vec{v}=-\frac{q}{c}\vec{H}
\]
equation (\ref{first madelung fluid}) can be rewritten as:
\begin{equation}\label{third madelung fluid}
 m\left(\frac{\partial \vec{v}}{\partial t}+(\vec{v}\cdot
  \nabla)\vec{v}\right)=q\textbf{E}+\frac{q}{c}\vec{v}\times\textbf{H}.
\end{equation}

The continuity equation is transformed into:
\begin{equation}\label{fourth madelung fluid}
  \frac{\partial \rho}{\partial t}+\grad\cdot(\rho \vec{v})=-\frac{\grad\cdot(\rho \grad \Phi)}{m},
\end{equation}
Therefore, the number of particles is not locally conserved.

Equations (\ref{third madelung fluid}) and (\ref{fourth madelung
fluid}) describe a flux of classical particles, under the action
of the electromagnetic field.

\section{Conclusions and Remarks}
If the ideas exposed in this paper are proven to be valid:
\begin{enumerate}
  \item The main assumption of the de Broglie-Bohm theory---that the local
impulse of quantum particles is given by the gradient of the phase
of the wave function---wont be accurate. [See eq.
(\ref{klein-gordon four-velocity}).]
  \item However, there will be still room for a classical interpretation of
quantum phenomena, in terms of particles moving along well defined
trajectories, under the action of the electromagnetic field.
\item The number of particles wont be locally conserved.
\item Given that a hidden variable has been introduced---after some considerations on electrodynamics---the
representation of physical systems through wave functions wont be
complete and, therefore, as foreseen by Einstein, Podolsky, and
Rosen, quantum mechanics wont be a complete theory of motion.
Actually, but for the restrictions imposed in the low-speed limit,
the function $\Phi$ could be equal to $\phi-S$ for any solution of
equation (\ref{relativistic hamilton-jacobi}).
\end{enumerate}

\end{document}